\documentclass[lettersize,journal]{IEEEtran}
\usepackage{amsmath,amsfonts}
\usepackage{algorithmic}
\usepackage{algorithm}
\usepackage{array}
\usepackage[caption=false,font=normalsize,labelfont=sf,textfont=sf]{subfig}
\usepackage{textcomp}
\usepackage{stfloats}
\usepackage{url}
\usepackage{verbatim}
\usepackage{graphicx}
\usepackage{bm}
\usepackage{cite}
\usepackage{booktabs}
\usepackage{multirow}
\usepackage{color}
\usepackage{makecell}
\usepackage{xcolor}
\usepackage{gensymb}
\usepackage{hyperref}
\hypersetup{hidelinks}
\usepackage{cleveref}
\usepackage{geometry}
\graphicspath{{imgs/}}
\begin{document}
\title{Coordinated Optimal Power Quality Management in Distribution Systems Using The Residual Capacity of Community IBRs}
\author{Tiantian Ji,~\IEEEmembership{Member,~IEEE},
        Pengfeng Lin,~\IEEEmembership{Member,~IEEE},
        Miao Zhu,~\IEEEmembership{Senior Member,~IEEE},
        Stephan M. Goetz,~\IEEEmembership{Senior Member,~IEEE},
        Ahmed Abu-Siada, ~\IEEEmembership{Senior Member,~IEEE},
        Syed Islam, ~\IEEEmembership{Fellow,~IEEE}
%
%
%
%
%
}



\maketitle
\begin{abstract} 
This letter proposes a network-wide coordinated optimization model to mitigate voltage unbalance (VU) by unleashing the remaining capacity of community inverter-based resources (IBRs). Existing single-sequence strategies ignore coupled capacity constraints and cause idle headroom. Meanwhile, they fail to harness the collective governance capabilities of community IBRs. To solve this discrepancy and exploit the unused potential, we developed a sequence-domain network model in dual commonly shared synchronous reference frames. Strict phase current and apparent power limits are formulated and convexified via polyhedral approximations. A quadratic objective function flexibly balances sequence capacity allocation. Simulation and experimental results validate the effectiveness of the proposed strategy.
\end{abstract}
\section{Introduction}
Voltage unbalance (VU) has emerged as a recurring power-quality issue in renewable‑dominant power grids. 
It typically manifests a decrease in the positive-sequence voltage and an increase in the negative-sequence voltage, which severely compromises system reliability, increases line losses, and limits the hosting capacity of the system \cite{Tiantian}.
 Traditional voltage-unbalance mitigation (VUM) strategies primarily rely on dedicated compensation devices, such as static var compensators \cite{STATCOM}.
  Although effective, the deployment of compensation devices inevitably entails additional hardware investment and operational expenditures.
  Moreover, prior to commissioning, such equipment must undergo grid permitting and compliance checks, which may prolong project implementation.

Meanwhile, most renewable energy resources are interfaced with the grid through electronic converters. The growing renewable sources lead to a large number of inverter-based resources (IBRs) in distribution networks. 
 Since IBRs are often not fully loaded, substantial capacity headroom is inherently available.
  Consequently, both academia and industry are increasingly turning to a promising solution: unleashing the remaining capacity of IBRs for VUM. 
  A straightforward approach uses negative-sequence voltage-droop control to generate compensation current references of IBRs \cite{droop1, droop2}. 
   Once the inverter controller detects a negative-sequence voltage, it  will calculate the negative-sequence current references based on the predefined droop coefficients.
 However, this local control scheme may generate infeasible current references under severe unbalanced conditions. Furthermore, it leads to suboptimal coordination among multiple distributed IBRs, which limits their overall potential.

 Several studies have attempted to maximize the capacity utilization of IBRs for VUM through optimization-based methods \cite{Camacho, Guo1, Guo2}.
 Camacho et al. derives the optimal current references under different grid impedance conditions \cite{Camacho}. However, when the active power output is constrained, the obtained solutions may become suboptimal.
Guo et al. study this limitation further \cite{Guo1,Guo2}; they focus on positive-sequence voltage support and negative-sequence mitigation. Both works visualise the current and active power constraints jointly in the current plane to identify the optimal operating point through geometric intersection. 
  Nevertheless, these single-objective approaches overlook the coupled capacity occupation when positive- and negative-sequence regulations are executed simultaneously. Additionally, while Guo et al. explore multi-IBR coordination, it relaxes the active power constraint to derive analytical solutions, which enforces only current limits \cite{Guo2}.
  
  To bridge these gaps, this letter proposes a coordinated optimization model for community IBRs to achieve network-wide VUM. 
  The main contributions include: 
  1) A sequence-domain network model within dual commonly shared reference frames, which effectively eliminates complex bus phase-angle variables. 
 2) Comprehensive IBR constraints that encompass phase current, sequence voltage, and apparent power limits. Furthermore, we thoroughly evaluate the tightness of these constraints and derive polyhedral approximations to reduce the nonlinear dimensionality.
 3) A quadratic objective function to flexibly allocate positive- and negative-sequence capacities.

\section{Sequence-Domain Network Modeling}
Consider a radial distribution network with multiple IBRs connected at different buses (Fig. \ref{Fig. distribution}). Bus 0 represents the upstream unbalanced grid and is modeled as an infinite bus. Voltage unbalance in the downstream network originates from Bus 0. Each IBR is modeled as a controlled current source that injects three-phase currents into the network.

\begin{figure}
	\setlength{\abovecaptionskip}{-3pt}
	\setlength{\belowcaptionskip}{-5pt}
	\centering
	\includegraphics[scale=1]{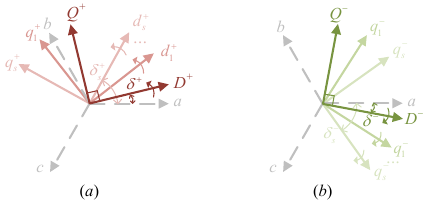}
	\caption{Reference frame transformation. The $D^+$- and $ D^-$-axes are initially aligned with phase $a$, with phases starting at zero.}
	\label{Fig. DQ frame}
\end{figure}

The analysis uses dual commonly shared synchronous rotating reference frames (Fig. \ref{Fig. DQ frame}): $DQ^+$ for the positive sequence and $DQ^-$ for the negative sequence. 
The dual reference frames circumvent the phase angle discrepancies between different local $dq$ frames. 
This letter uses superscripts “$^+$” and “$^-$” to respectively denote positive- and negative-sequence electrical quantities.
In a distribution network with $m$ buses, the positive- and negative- sequences power flow equations are respectively
\begin{equation}\label{power flow}
	\left\{
	\begin{aligned}
		\vec{\boldsymbol{V}}^+ &= \vec{V}_{0}^+\mathbf{1}_m 
		+ \boldsymbol{Z}_{\mathrm{net}}\left(\vec{\boldsymbol{I}}^+ - \boldsymbol{Y}_L \vec{\boldsymbol{V}}^+\right)\!,\\
		\vec{\boldsymbol{V}}^- &= \vec{V}_{0}^-\mathbf{1}_m 
		+ \boldsymbol{Z}^*_{\mathrm{net}}\left(\vec{\boldsymbol{I}}^- - \boldsymbol{Y}^*_L \vec{\boldsymbol{V}}^-\right)\!,
	\end{aligned}
	\right.	
\end{equation}
where $\vec{\boldsymbol{V}}^\phi = [\vec{V}_{1}^\phi, \dots, \vec{V}_{m}^\phi]^T$ and
$\vec{\boldsymbol{I}}^\phi = [\vec{I}_{1}^\phi, \dots, \vec{I}_{m}^\phi]^T$ ($\phi \in \{+,-\}$) respectively denote
the nodal voltage and current injection vectors.
Specifically, the $i$-th elements are defined as phasors $\vec{V}_{i}^\phi = V_{i}^\phi \angle\varphi_{i}^\phi$
and $\vec{I}_{i}^\phi = I_{i}^\phi \angle\theta_{i}^\phi$.
The term $\vec{V}_{0}^\phi = V_{0}^\phi \angle\varphi_{0}^\phi$ denotes the slack bus voltage, which is broadcast to all buses via the all-ones vector $\boldsymbol{1}_m$.
$\boldsymbol{Y}_L = \mathrm{diag}(Y_{L1}, \dots, Y_{Lm})$
denote the load admittance matrix. The superscript “$^{*}$” denotes the complex conjugate operator.
Finally, $\boldsymbol{Z}_{\rm{net}} = \left[ Z_{ij} \right]_{m \times m}$ is the network impedance matrix, with entries calculated per  
\begin{equation} \label{Znet}
	Z_{ii} = \sum_{l \in \mathcal{L}_i} z_l,\:Z_{ij} = \sum_{l \in \mathcal{L}_i \cap \mathcal{L}_j} z_l,
\end{equation}
where $z_l$ is the impedance of line $l$. $\mathcal{L}_{i}$ is the set of lines on the unique path from the bus 0 to bus $i$, $i \in \mathcal{M}$, $j \in \mathcal{M}$, $\mathcal{M}:=\{1, \dots, m\}$.

Let $(\boldsymbol{I}_m + \boldsymbol{Z}_{\mathrm{net}} \boldsymbol{Y}_L)^{-1} \boldsymbol{Z}_{\mathrm{net}}=\boldsymbol{H} \boldsymbol{Z}_{\mathrm{net}}=\boldsymbol{Z}^{\mathrm{eq}}$, rearranging \eqref{power flow} yields the following equivalent form

\begin{equation}
	\left\{
	\begin{aligned}
		\vec{\boldsymbol{V}}^+ &= \boldsymbol{H} \vec{V}_{0}^+ \mathbf{1}_m + \boldsymbol{Z}^{\mathrm{eq}} \vec{\boldsymbol{I}}^+\!, \\
		\vec{\boldsymbol{V}}^- &= \boldsymbol{H}^* \vec{V}_{0}^- \boldsymbol{1}_m + ( \boldsymbol{Z}^{\mathrm{eq}})^* \vec{\boldsymbol{I}}^-\!.
	\end{aligned}
	\right.
\end{equation}

The nodal voltage at bus $i$ is derived as
\begin{equation}\label{Vsp/Vsn}
	\left\{
	\begin{aligned}
		\vec{V}_{i}^+ &= \sum_{j=1}^{m}H_{ij} \vec{V}_{0}^+ + \sum_{j=1}^{m} Z^{\rm eq}_{ij}\vec{I}_{j}^+\!,\\
		\vec{V}_{i}^- &= \sum_{j=1}^{m}H_{ij}^* \vec{V}_{0}^- + \sum_{j=1}^{m} (Z^{\rm eq}_{ij})^* \vec{I}_{j}^-\!.
	\end{aligned}
	\right.
\end{equation}
\section{Proposed Coordinated Optimization Formulation}
\subsection{Operational Constraints}
\begin{figure}
	\setlength{\abovecaptionskip}{-3pt}
	\setlength{\belowcaptionskip}{-5pt}
	\centering
	\includegraphics[scale=0.2]{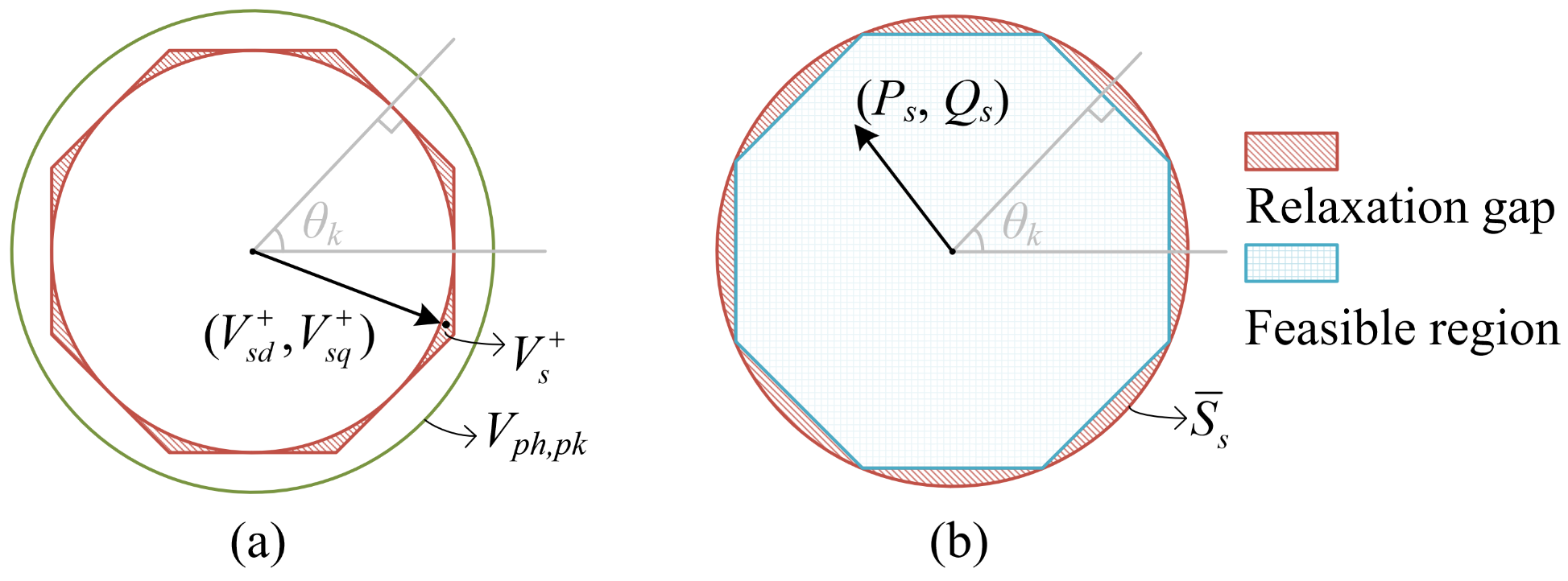}
	\caption{Illustration of the polyhedral relaxation and approximation: (a) Circumscribed $n$-sided polygon. (b) Inscribed $n$-sided polygon. Here, $\theta_k = \frac{2(k-1)\pi}{n}$ represents the normal direction angle of the $k$-th side.}
	\label{Fig. polyhedral}
\end{figure}
\textit{1) Nodal Voltage Constraints:}
Projecting \eqref{Vsp/Vsn} onto the $DQ^+$ and $DQ^-$ frames yields \eqref{dq}.
\begin{figure*}
	\begin{equation}\label{dq}
		\small 
		\begin{aligned}
			V_{id}^+ &= V_{0}^+ \sum_{j=1}^{m} (G_{ij}\cos\varphi_{0}^+ \!-\! B_{ij}\sin\varphi_{0}^+) + \sum_{j=1}^{m} (R^{\text{eq}}_{ij} I_{jd}^+ \!-\! X^{\text{eq}}_{ij} I_{jq}^+), \hspace{-1em}&
			V_{iq}^+ &= V_{0}^+ \sum_{j=1}^{m} (G_{ij}\sin\varphi_{0}^+ + B_{ij}\cos\varphi_{0}^+) \!+\! \sum_{j=1}^{m} (R^{\text{eq}}_{ij} I_{jq}^+ + X^{\text{eq}}_{ij} I_{jd}^+), \\
			V_{id}^- &= V_{0}^- \sum_{j=1}^{m} (G_{ij}\cos\varphi_{0}^- \!+\! B_{ij}\sin\varphi_{0}^-) \!+\! \sum_{j=1}^{m} (R^{\text{eq}}_{ij} I_{jd}^- \!+\! X^{\text{eq}}_{ij} I_{jq}^-), \hspace{-1em}&
			V_{iq}^- &= V_{0}^- \sum_{j=1}^{m} (G_{ij}\sin\varphi_{0}^- - B_{ij}\cos\varphi_{0}^-) \!+\! \sum_{j=1}^{m} (R^{\text{eq}}_{ij} I_{jq}^- \!-\! X^{\text{eq}}_{ij} I_{jd}^-).
		\end{aligned}
	\end{equation}
   	\hrulefill
\end{figure*}
In \eqref{dq}, $G_{ij}$ and $B_{ij}$ are the real and imaginary parts of $H_{ij}$; $R_{ij}$ and $X_{ij}$ are the real and imaginary parts of $Z_{ij}$.
Subsequently, the voltage magnitude constraints are formulated as \eqref{relax_amp} by summing the squares of the $d$- and $q$-axis components and relaxing the equalities:
\begin{subequations}\label{relax_amp}
	\begin{align}
		(V_{i}^+)^2 &\ge (V_{id}^+)^2 + (V_{iq}^+)^2, i \in \mathcal{M} \label{relax_amp_p}
		\\
		(V_{i}^-)^2 &\ge (V_{id}^-)^2 + (V_{iq}^-)^2, i \in \mathcal{M} \label{relax_amp_n}
	\end{align}
\end{subequations}

In addition, $V_{i}^+$ and $V_{i}^-$ should be non-negative and not exceed the rated peak phase voltage $V_\mathrm{ph,pk}$, as expressed in \eqref{eq:voltage_bounds}
\begin{equation} \label{eq:voltage_bounds}
	0 \le V_{i}^+ \le V_\mathrm{ph,pk}, \quad 0 \le V_{i}^- \le V_\mathrm{ph,pk}.
\end{equation}

\textit{2) Current Constraints:} Let $\mathcal{S}$ denote the set of buses equipped with IBRs. The IBR connected to bus $s$ is referred to as IBR-$s$, $s \in \mathcal{S}$. In the $DQ^+$ and $DQ^-$ frames, the time-domain expression of the phase A current output by IBR-$s$ is given as
\begin{equation}
	\begin{aligned}
		{i_{sa}} &= i_{sa}^+ + i_{sa}^-\\
		&=(I_{sd}^+ \cos \delta^+ - I_{sq}^+ \sin \delta^+) + (I_{sd}^- \cos \delta^+ + I_{sq}^- \sin \delta^+).
	\end{aligned}
\end{equation}

Based on Cauchy's inequality, eliminating $\delta^+$ yields
\begin{equation}\label{ia}
	\small
	\begin{array}{c}
		{i_{sa}}
		\le \sqrt {{{({I_{sd}^+} + {I_{sd}^-})}^2} + {{({I_{sq}^+} - {I_{sq}^-})}^2}}  \le {{\overline I}_s},
	\end{array}
\end{equation}
where $\bar I_s$ denotes the steady–state phase current magnitude limit.

Similarly, the current inequalities for phases B and C can be obtained as 
\begin{equation}\label{ib}
	\begin{aligned}
		i_{sb} \le 
		\Big[ &\big(I_{sd}^+ + I_{sd}^-\cos\tfrac{4\pi}{3} + I_{sq}^-\sin\tfrac{4\pi}{3}\big)^2 \\
		&+ \big(I_{sq}^+ + I_{sd}^-\sin\tfrac{4\pi}{3} - I_{sq}^-\cos\tfrac{4\pi}{3}\big)^2
		\Big]^{\tfrac{1}{2}}
		\le \overline I_s,
	\end{aligned}
\end{equation}
\begin{equation}\label{ic}
	\begin{aligned}
		i_{sc} \le 
		\Big[ &\big(I_{sd}^+\cos\tfrac{4\pi}{3} + I_{sd}^- - I_{sq}^+\sin\tfrac{4\pi}{3}\big)^2 \\
		&+ \big(I_{sq}^+\cos\tfrac{4\pi}{3} + I_{sd}^+\sin\tfrac{4\pi}{3} - I_{sq}^-\big)^2
		\Big]^{\tfrac{1}{2}}
		\le \overline I_s.
	\end{aligned}
\end{equation}

\textit{3) Power Constraints:} 
The apparent power output of IBR-$s$ is constrained by
\begin{equation}\label{S}
	P_s^2 + Q_s^2 \le \overline S_s^2, s \in \mathcal{S}
\end{equation}
where $\overline S_s$ denotes the rated apparent power.
The expressions for $P_s$ and $Q_s$ are
\begin{subequations}\label{PQ}
	\begin{align}
		P_s &= 1.5(V_{sd}^+ I_{sd}^+ + V_{sq}^+ I_{sq}^+ + V_{sd}^- I_{sd}^- + V_{sq}^- I_{sq}^-), \\
		Q_s &= 1.5(V_{sq}^+ I_{sd}^+ - V_{sd}^+ I_{sq}^+ + V_{sq}^- I_{sd}^- - V_{sd}^- I_{sq}^-).
	\end{align}
\end{subequations}

Moreover, the IBR should satisfy the minimum power supply requirements or power absorption limits, i.e.,
\begin{equation}\label{P/Q}
	\underline{P}_s \le P_s, 	\quad \underline{Q}_s \le Q_s.
\end{equation}

\subsection{Polygonal Approximation and Tightening}
\textit{1) Tightening of Positive-Sequence Voltage:}
Since the optimization objective is to maximize $V_{i}^+$\!, \eqref{relax_amp_p} fails to be tight and drives $V_{i}^+$ toward its theoretical upper bound $V_\mathrm{ph,pk}$ rather than its physical value. To rectify this, the original constraint \eqref{relax_amp_p} is augmented with a circumscribed $n$-sided polyhedral envelope (Fig. \ref{Fig. polyhedral}(a)).
Hence, $V_{i}^+$ is jointly constrained by \eqref{relax_amp_p} and \eqref{poly_tight and binary}.

\begin{subequations} \label{poly_tight and binary}
	\begin{align}
		V_{i}^+ \cos(\frac{\pi}{n})  & \le V_{id}^+\cos\theta_k + V_{iq}^+\sin\theta_k+M(1-x_i^k)
		\\
		\sum_{k=1}^n x_i^k & \le 1, i \in \mathcal{M}
	\end{align}
\end{subequations}
where $k=1, \dots, n$. $M$ is a big value which is set as the rated voltage of a given system. $x_i^k \in \{0,1\}$ is a binary variable for the $k$-th side of circumscribed polyhedral at bus-$i$.

\textit{2) Approximation of Power Constraints:}
To handle the strong non-linearity of the power constraint \eqref{S}, an inscribed $n$-sided polygon is employed to approximate the circular feasible region, as illustrated in Fig. \ref{Fig. polyhedral}(b). Consequently, \eqref{S} is replaced by the following $n$ linear inequalities:
\begin{equation} \label{poly_tight2}
	\overline S_s  \cos\left(\frac{\pi}{n}\right) \ge P_{s}\cos\theta_k + Q_{s}\sin\theta_k , k=1, \dots, n.
\end{equation}

Both of the polygonal constraints above consistently use $n=8$ to strike a favorable balance between the relaxation gap and computational efficiency. 
\subsection{Objective Function}
Although the voltage unbalance factor (VUF), defined as $V^- / V^+$, is a usual metric for quantifying asymmetry, it is unsuitable as a direct optimization objective.
Once the negative-sequence voltage is fully eliminated, the ratio becomes zero regardless of the positive-sequence voltage magnitude.
Hence, this letter adopts a quadratic objective function $J$ that simultaneously minimizes the negative-sequence voltage and constrains the positive-sequence voltage close to its nominal value.
In summary, the optimization model is formulated as
\begin{equation} \label{opt_problem}
	\begin{aligned}
		\min &\quad J=\sum_{i \in \mathcal{N}} \left( \frac{V_i^-}{V_{\mathrm{ph,pk}}} \right)^2
		+  \lambda \left( \frac{V_i^+}{V_{\mathrm{ph,pk}}} - 1 \right)^2, \\
		\mathrm{s.t.} &\quad \eqref{relax_amp}-\eqref{ic},\eqref{P/Q}-\eqref{poly_tight2},
	\end{aligned}
\end{equation}
where $\lambda$ acts as a weighting factor to adjust the capacity allocation ratio between the positive- and negative-sequence regulations. This letter sets $\lambda$ to 1. The solution of \eqref{opt_problem} delivers the optimal current references of all IBRs under the $DQ$ frame. 
These references are then transformed back to the local $dq$ frame of each inverter for implementation. $\mathcal{N}$ is a set wherein each component is the index of bus of interest as per system operators. $\mathcal{N}$ can be either  $\mathcal{M}$ or  $\mathcal{S}$.

\section{Case Studies}
\subsection{Simulation Verification}
\begin{figure}[htbp]
	\setlength{\abovecaptionskip}{-3pt}
	\setlength{\belowcaptionskip}{-5pt}
	\centering
	\includegraphics[scale=1.16]{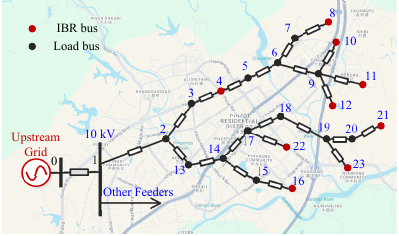}
	\caption{Topology of the modified 23-bus distribution network. The network dataset and main code used in simulation are available at \url{https://github.com/missinglpf/Coordinated_Optimal_Power_Quality_Management.git}.}
	\label{Fig. distribution}
\end{figure}

The simulation topology shown in Fig. \ref{Fig. distribution} is modified from an actual regional grid in Longgang District, Shenzhen, China. 
The optimization performance of the proposed coordinated strategy (S3) is evaluated and compared with two optimization methods:  the positive-sequence voltage support strategy (S1) and the negative-sequence voltage attenuation strategy (S2). 
S1 and S2 are derived from the original single-IBR strategies presented in \cite{Guo2} and \cite{Guo1}, respectively, and they are refined to accommodate the coordination of community IBRs.

 Two comparative cases are considered:
\begin{itemize}
	\item \textbf{Case 1 (Moderate VU):} $\vec{V}_{0}^+ = 0.8 \,\angle\, 0^\circ$~p.u., $\vec{V}_{0}^- = 0.1 \,\angle\, -90^\circ$~p.u.
	\item \textbf{Case 2 (Severe VU):} $\vec{V}_{0}^+ = 0.6 \,\angle\, 0^\circ$~p.u., $\vec{V}_{0}^- = 0.4 \,\angle\, -30^\circ$~p.u.
\end{itemize}

\begin{figure}[htbp]
	\setlength{\abovecaptionskip}{-3pt}
	\setlength{\belowcaptionskip}{-5pt}
	\centering
	\includegraphics[scale=1]{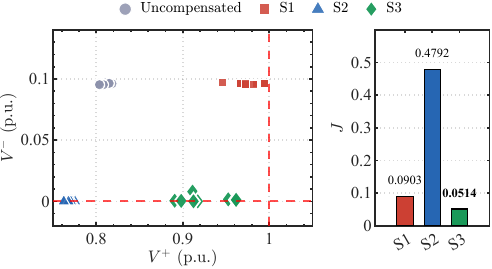}
	\caption{Performance comparison of different strategies (Case 1). Left: Sequence voltage distribution of IBR buses. Right: Optimization objective values.}
	\label{Fig. case1}
\end{figure}

\begin{figure}[htbp]
	\setlength{\abovecaptionskip}{-3pt}
	\setlength{\belowcaptionskip}{-5pt}
	\centering
	\includegraphics[scale=1]{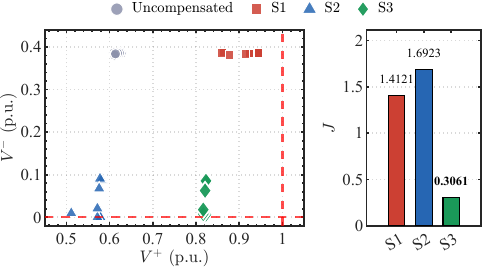}
	\caption{Performance comparison of different strategies (Case 2). Left: Sequence voltage distribution of IBR buses. Right: Optimization objective values.}
	\label{Fig. case2}
\end{figure}

Fig.~\ref{Fig. case1} compares the control performances of the different strategies under Case 1.
As observed in the left scatter plot, S1 and S2 perform voltage regulation solely from either the positive- or negative-sequence perspective. 
Consequently, the inverter cannot exploit its remaining capacity effectively. For instance, although S2 successfully regulates the negative-sequence voltages to zero, its residual capacity is left idle.
Moreover, they ignore the fact that both sequence regulations simultaneously occupy the total inverter capacity.
 In contrast, the proposed S3 explicitly accounts for the coupled capacity occupation. By regulating power quality from both sequence dimensions, S3 fully exploits the available headroom.
Quantitatively, the right bar chart demonstrates that S3 achieves the minimum objective value of 0.0514, which significantly outperforms S1 and S2.
Fig.~\ref{Fig. case2} further illustrates the control performance under \textit{Severe VU} (Case 2). As the VU becomes more severe, the required compensation capacity significantly increases, and the limitations of single-sequence strategies become more pronounced. Specifically, S2 commits its limited capacity to eliminate the severe negative-sequence voltage. In consequence, it leaves insufficient capacity for positive-sequence voltage support, which causes $V^+$ to plummet below 0.6 p.u.
In comparison, the proposed S3 strikes a balance in positive- and negative-sequence capacity allocation and maintains the minimum objective value.


\subsection{Experimental Verification}
An experimental hardware platform validated the feasibility of the proposed strategy (Fig. \ref{Fig. Experimental platform}).
 The setup comprises two identical IBRs, each powered by a 400 Vdc source. 
A dSpace1202 controller with a switching frequency of 10 kHz executes the overall control method. 
The remaining system settings are directly annotated in Fig. \ref{Fig. Experimental platform}. 
An unbalanced grid condition is emulated with $\vec{V}_0^+ = 0.8 \angle 90^\circ$ p.u. and $\vec{V}_0^- = 0.1 \angle 15^\circ$ p.u. 
 Table \ref{tab:sequence_currents} lists the sequence current references for IBR-1 and IBR-2, calculated by the proposed strategy.
\begin{table}[htbp]
	\setlength{\abovecaptionskip}{-2pt}
	\setlength{\belowcaptionskip}{-2pt}
	\vspace{-0.2em}
	\centering
	\caption{Optimal Sequence Currents}
	\label{tab:sequence_currents}
	\begin{tabular}{lcccc}
		\toprule
		\textbf{Unit} & $I_{d}^+$ (A) & $I_{q}^+$ (A) & $I_{d}^-$ (A) & $I_{q}^-$ (A) \\ 
		\midrule
		\textbf{IBR-1} & 3.20 & 4.14 & -4.36 & -4.16 \\ 
		\textbf{IBR-2} & 5.68 & 4.75 & -2.60 & -2.29 \\ 
		\bottomrule
	\end{tabular}
	\vspace{-0.5em}
\end{table}
\begin{figure}[htbp]
	\setlength{\abovecaptionskip}{-3pt}
	\setlength{\belowcaptionskip}{-5pt}
	\centering
	\includegraphics[scale=1]{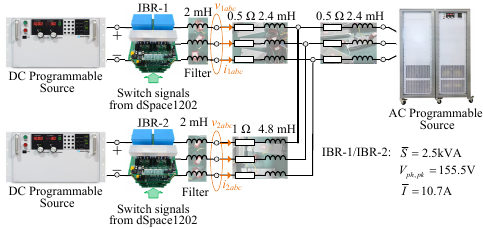}
	\caption{Experimental platform.}
	\label{Fig. Experimental platform}
\end{figure}

\begin{figure}[htbp]
	\setlength{\abovecaptionskip}{-3pt}
	\setlength{\belowcaptionskip}{-5pt}
	\centering
	\includegraphics[scale=1.06]{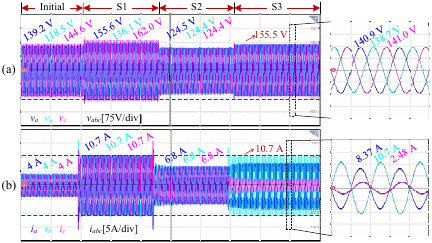}
	\caption{Experimental results of the IBR-1 under various control schemes: (a) Three-phase terminal voltages $v_{1abc}$, (b) three-phase output currents $i_{1abc}$.}
	\label{Fig. IBR-1.}
\end{figure}

\begin{figure}[htbp]
	\setlength{\abovecaptionskip}{-3pt}
	\setlength{\belowcaptionskip}{-5pt}
	\centering
	\includegraphics[scale=1.06]{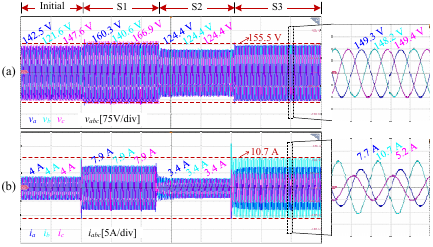}
	\caption{Experimental results of the IBR-2 under various control schemes: (a) Three-phase terminal voltages $v_{2abc}$, (b) three-phase output currents $i_{2abc}$.}
	\label{Fig. IBR-2.}
\end{figure}

Fig. \ref{Fig. IBR-1.} illustrates the output voltage and current waveforms of IBR-1 under different optimization strategies. In the Initial stage, a significant unbalanced voltage exists at the IBR’s output terminals. Upon entering the S1 stage, the positive-sequence voltage is effectively enhanced; however, since the negative-sequence voltage remains unadjusted, the phase A voltage exceeds the operational limit. In the S2 stage, the negative-sequence voltage is successfully mitigated, yet substantial IBR capacity remains under-utilized as the output currents are well below their ratings. Under the proposed S3 strategy, the system prioritizes minimizing the negative-sequence voltage while it simultaneously boosts the positive-sequence level. Notably, the current of phase B  reach its predefined limit of 10.7 A. 
Also, Fig. \ref{Fig. IBR-2.} shows similar regulation characteristics for IBR-2. These results demonstrate that the proposed strategy can fully exploit the IBR's power capacity while strictly adhering to hardware security constraints.

\section{Conclusion}
This letter has proposed a network-wide coordinated optimization strategy to mitigate voltage unbalance by fully exploiting the remaining capacity of distributed IBRs.
The core contribution lies in the formulation of strict phase current and apparent power constraints in tailed frames and a polygonal substitution approach, which avoid complex phase-angle variables and ensure a tractable optimization model.  
Comprehensive simulations and experiments validated that the proposed strategy fully unleashes the remaining capacity of IBRs and effectively achieve optimal power quality management.
\bibliographystyle{IEEEtran}
\bibliography{reference.bib}
\vspace{11pt}
\vfill

\end{document}